\documentstyle[twocolumn,aps,epsf,floats]{revtex}
\begin{document}

\twocolumn[\hsize\textwidth\columnwidth\hsize\csname %
@twocolumnfalse\endcsname

\title{Effect of next-nearest neighbor hopping on the spin dynamics in antiferromagnets}
\author{Dirk K. Morr}
\address{University of Illinois at Urbana-Champaign, 
Loomis Laboratory of Physics\\
1110 W. Green St., Urbana, Il, 61801}
\date{\today}
\maketitle
\begin{abstract}
Recently, inelastic neutron scattering (INS) experiments on the insulating parent compounds of high-$T_c$ materials were analyzed to extract the value of the superexchange constant $J$. Starting point of the analysis was the nearest-neighbor Heisenberg model.  Motivated by recent ARPES experiments, 
we consider the effects of a  next-nearest neighbor 
hopping, $t^\prime$ in the strong coupling limit of the spin-density wave formalism, where it leads to an 
antiferromagnetic exchange $J^\prime>0$ between next-nearest neighbor spins. We show that a finite $J^\prime$ (a) changes
the spin-wave velocity and thus the value of $J$ extracted from the INS data, yielding almost identical values for $J$ in La$_2$CuO$_4$ and YBa$_2$Cu$_3$O$_{6.15}$, and (b) leads to a considerable change in the high-frequency spin-dynamics.  Finally, we calculate the $1/S$ quantum corrections in the presence of $J^\prime$ and discuss their relevance for the INS data. 
\end{abstract}
 ]

\narrowtext

The strong evidence for antiferromagnetic fluctuations as the source of the pairing mechanism in the high-$T_c$ materials \cite{Pin97,Sca95} makes it important to understand the variation of the energy scale $J$ of these fluctuations in different compounds. The materials for which $J$ can be easily determined are the insulating parent compounds of the high-$T_c$ materials. Here, the energy scale $J$ of the spin-wave mode can be extracted from Raman \cite{Lyo88} or inelastic neutron scattering (INS) experiments\cite{Hay97}. Recently, INS experiments \cite{Hay97,Bou97} on the insulating parent compounds La$_2$CuO$_4$, Nd$_2$CuO$_4$, Pr$_2$CuO$_4$ and YBa$_2$Cu$_3$O$_{6.15}$, which possess an antiferromagnetic ground state, were analyzed using the Heisenberg model with nearest-neighbor exchange $J$. Combining the experimentally measured spin-wave velocity $c_{sw}$ with the theoretical expression $c_{sw}=\sqrt{2}J$ of the Heisenberg model,  values of $J$ were obtained which range from $J \approx 125$ meV in YBa$_2$Cu$_3$O$_{6.15}$ to $J \approx 160$ meV in Nd$_2$CuO$_4$.
However, various theoretical studies \cite{Duf95,Chu95} of angle-resolved photoemission (ARPES) experiments \cite{Wells} as well as bandstructure calculations \cite{Hyb90} suggest, that the experimentally measured fermionic dispersion provides evidence for a finite next-nearest neighbor hopping, $t^\prime$, with $t^\prime/t\approx-0.25$ for La$_2$CuO$_4$ and $t^\prime/t\approx-0.45$ for YBa$_2$Cu$_3$O$_{6.15}$.
In the strong coupling limit of the insulating parent compounds, $t^\prime$ leads to an antiferromagnetic coupling $J^\prime>0$  between next-nearest neighbors spins which changes $c_{sw}$ and in turn the value of  $J$ extracted from the INS data.

In this communication we investigate the effect of  $J^\prime$ on the  spin dynamics in single and double-layer compounds. We show that  $J^\prime$ not only changes the spin-wave velocity but also the  spin-dynamics at  higher frequencies. We reanalyze the INS data and demonstrate that $J$ extracted in the presence of a finite $J^\prime$ is actually larger than previously assumed \cite{Hay97}. In particular, we obtain almost identical values for $J$ in La$_2$CuO$_4$ and YBa$_2$Cu$_3$O$_{6.15}$. Furthermore, we compute  $1/S$ (quantum) corrections \cite{Sin89,Iga92} for various physical quantities. These corrections increase with increasing  $J^\prime$, which is expected since $J^\prime>0$ frustrates the system and should thus lead to an increase of quantum fluctuations. Finally, we discuss our results in the context of the experimentally observed strong renormalization of the transverse spin susceptibility.

We study the effects of $t^\prime$ on the spin dynamics by computing the transverse susceptibility within the spin-density wave (SDW)  formalism at $T=0$. 
Since this formalism is thoroughly described in the literature \cite{SWZ}, we will outline it only briefly. For simplicity we first consider the case of a single-layer compound. The starting point for the description of the CuO$_2$ planes in the insulating parent compounds is the effective 2D one-band Hubbard model \cite{Sca95,Hubbard} at half filling, given by
\begin{equation}
{\cal H} = - \sum_{i,j} t_{i,j} c^{\dagger}_{i,\alpha} c_{j,\alpha} 
+ U \sum_i c^{\dagger}_{i,\uparrow} c_{i,\uparrow} c^{\dagger}_{i,\downarrow} 
c_{i,\downarrow} \ .
\label{hub}
\end{equation}
Here $\alpha$ is the spin index and $t_{i,j}$ is the hopping
integral which we assume to act between nearest and next-nearest neighbors
($t$ and $t^{\prime}$, respectively). In the strong coupling limit, $t \ll U$ the ground state of the Hubbard model is antiferromagnetically ordered, and one can use the antiferromagnetic long range order parameter $\langle S_z \rangle = \sum_k \langle c^{\dagger}_k c_{k+Q} \rangle$ to decouple the interaction term in Eq.(\ref{hub}). Diagonalizing then the quadratic form by means of a unitary transformation
one obtains two electronic bands for the conduction and valence fermions 
whose mean-field energy dispersions are given by
\begin{equation}
E^{c,v}_ k = \pm \sqrt{(\epsilon^{-}_ k)^2 + \Delta^2}+
 \epsilon^{+}_k \ ,
\label{f_disp}
\end{equation}
where $\epsilon^{\pm}_k = (\epsilon_k \pm \epsilon_{k+Q}) /2,  \Delta = U \langle S_z \rangle$, and
\begin{equation}
\epsilon_ k =-2t (\cos k_x + \cos k_y) - 4 t^{\prime} \cos k_x \cos k_y - \mu \ .
\end{equation}
Here $E^{c,v}_ k$ is the dispersion of
the conduction and valence fermions, 
respectively, $\epsilon_k$ is the dispersion of free fermions, and 
 $\mu$ is the chemical potential.

Next we calculate the transverse susceptibility which is defined by
\begin{equation}
\chi^{+-}({\bf q},{\bf q^\prime} ,t) = i \langle T S^+_{\bf q}(t) S^-_{\bf -q^\prime}(0) \rangle \ ,
\label{susc}
\end{equation}
where the spin operators are given by
\begin{equation}
{\bf S_q} = { 1 \over 2} \sum_k c^\dagger_{ {\bf k} + {\bf q}, \alpha} 
{\bf \sigma}_{\alpha,\beta} c_{ {\bf k}, \beta}  \ ,
\end{equation}
and ${\bf \sigma}$ are the Pauli matrices.  Within the RPA approximation $\chi^{+-}$ is given by an infinite series of fermionic bubble diagrams \cite{SWZ}, with fermions obeying the dispersion given in Eq.(\ref{f_disp}). 
Since the susceptibilities with zero momentum transfer (i.e. ${\bf q^\prime=q}$) and with momentum transfer ${\bf Q}=(\pi,\pi)$ (i.e. ${\bf q^\prime=q+Q}$) are finite, the evaluation of $\chi^{+-}$ reduces to a $(2 \times2)$ algebraic equation. One obtains after some tedious algebra (for S=$1/2$)
\begin{eqnarray}
& &\chi^{+-}({\bf q},{\bf q},\omega) = {-2J \Big[ 1 - \nu_{\bf q} - {J^\prime \over J} \Big( 1-cos(q_x)cos(q_y) \Big) \Big] \over \omega^2 - \omega^2({\bf q}) + i\delta }, \nonumber \\
& & \chi^{+-}({\bf q},{\bf q+Q},\omega) = -  { \omega \over \omega^2 - \omega^2({\bf q}) + i\delta } , 
\label{chiQ}
\end{eqnarray} 
where
$J= 4t^2 / U$,  $J^\prime=4(t^\prime)^2/ U$ are the superexchange constants and $\nu_{\bf q} = ( cos(q_x)+cos(q_y) )/2$.  The magnon dispersion $\omega({\bf q})$ is given by
\begin{equation}
\omega({\bf q}) = 2J \sqrt{\Big[1- {J^\prime \over J} ( 1-cos(q_x)cos(q_y)) \Big]^2 - \nu_{\bf q}^2 } \ .
\label{disp}
\end{equation}
A comparison of Eq.(\ref{disp}) with the mean-field magnon dispersion of the Heisenberg model with next-nearest neighbor exchange $J^\prime$ shows,  as expected, that in the strong coupling limit $t^\prime$ leads to an antiferromagnetic coupling $J^\prime>0$. 
 
From Eq.(\ref{chiQ}) we see that the terms $\sim J^\prime$ vanish in the limit  ${\bf q}=0$, which means that the static, uniform susceptibility 
$\chi_\perp=1/8J$ is not affected by $J^\prime$.
In contrast, $J^\prime$ changes the spin-wave velocity $c_{sw}$, 
which is  given by
\begin{equation}
c_{sw} = \sqrt{2} J \sqrt{1- 2 {J^\prime \over J} }  \ .
\label{ren}
\end{equation}
Using the hydrodynamical relation $\rho_s=c_{sw}^2 \chi_\perp$ we obtain for the spin stiffness $\rho_s=J(1- 2 J^\prime/J)/4$ which was previously reported in Ref.\cite{Chu95}.

\begin{figure} [t]
\begin{center}
\leavevmode
\epsfxsize=7.5cm
\epsffile{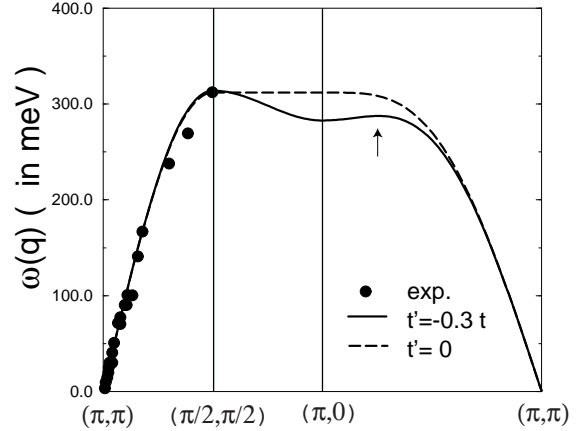}
\end{center}
\caption{Fit to the experimental magnon dispersion for 
$t^\prime=-0.3 t$ (solid line) and $t^\prime=-0.0$ (dashed line).
The experimental data are taken from Ref.\protect\cite{Hay91}.}
\label{cuts}
\end{figure}

We now discuss our results in the light of the available INS data. In Fig.~\ref{cuts} we present a fit of the magnon dispersion Eq.(\ref{disp}) for $t^\prime=-0.3t$ (solid line) to the experimental data on La$_2$CuO$_4$ from Ref.~\cite{Hay91}. 
This fit yields $J= 166 (\pm 6)$ meV, a value approximately $ 6 \%$ larger than in Ref.~\cite{Hay97}, and $J^\prime = 15 \pm 1$ meV.
For comparison, we also present a fit of the magnon dispersion with 
$t^\prime=0$ (dashed line). Note, that only the functional form of the high-frequency part of the dispersion is changed by $J^\prime$. Here, $J^\prime$ removes the degeneracy along the boundary of the magnetic Brillouin zone (MBZ) ($(0, \pi)$ to $(\pi, 0)$ direction), and gives rise to a local maximum  along the $(0,\pi)$ to $(\pi,\pi)$ direction (indicated by an arrow). It is, however, questionable whether INS experiments are sensitive enough to detect this difference  since the effect is only of the order of $10 \%$, and thus might very well lie within the experimental errorbars. 

\begin{figure} [t]
\begin{center}
\leavevmode
\epsfxsize=7.5cm
\epsffile{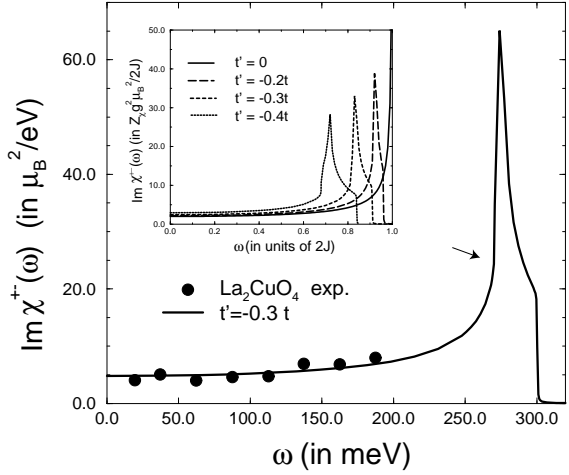}
\end{center}
\caption{Fit of the local susceptibility with $t^\prime=-0.3 t$ to the experimental data from Ref. \protect\cite{Hay97}. The inset shows the systematic evolution of $\chi''(\omega)$ with increasing $t^\prime/t$}
\label{chidp_om}
\end{figure}

Next, we analyze the local, or q-integrated imaginary part of the susceptibility, which is defined by
\begin{equation}
\chi''(\omega) = {1 \over 4 \pi^2 } \int d^2q \ Im \Big\{ 
\chi^{+-}({\bf q},{\bf q}, \omega) \Big\} \ .
\label{chi_om}
\end{equation}
In the low-frequency limit, we obtain
\begin{equation}
\chi''(\omega) = { 1 \over \sqrt{2} c_{sw} \sqrt{1-2 J^\prime/J} } \Bigg[ 1+
O\Big( \omega^2 \Big) \Bigg] 
\label{low_f1}
\end{equation}
Note that on the mean-field level,  $J^\prime$ increases the overall scale of  $\chi''(\omega)$ (this is opposite to the effect of quantum corrections, as we will discuss later).  In Fig. \ref{chidp_om} we present a fit of $\chi''(\omega)$ with $t^\prime=-0.3$ to the experimental data from Ref. \cite{Hay97}. For this comparison we multiplied Eqs.(\ref{chiQ}) and (\ref{low_f1}) with $Z_d g^2\mu_B^2$, where $g$ is the Land\'{e}-factor, $\mu_B$ is the Bohr magneton and $Z_d$ represents the overall renormalization due to quantum fluctuations \cite{Sin89,Iga92}. The inset shows the systematic evolution of  $\chi''(\omega)$ with increasing $t^\prime/t$.  
For $t^\prime=0$, $\chi''(\omega)$ is logarithmically divergent at the maximum magnon energy $\omega_{max}$ due to the degeneracy of the dispersion along the boundary of the MBZ. A finite $t^\prime$ removes this degeneracy (see Fig.~\ref{cuts}) and transforms the divergence into a maximum which shifts to lower energies with increasing $t^\prime/t$.
We can identify three characteristic features of $\chi''(\omega)$, which in addition to $c_{sw}$, provide enough information to extract $J^\prime$ from the experimental data. These features are the position of the maximum, the cut-off energy, and the position of the jump, denoted by an arrow in Fig.~\ref{chidp_om}.
The cut-off energy is obviously given by the maximum magnon energy $\omega_{max} = 2J - 2J^\prime$, but for small $J^\prime$ yields the same information as $c_{sw}$. To determine the position of the jump and the peak we observe that at higher frequencies $\chi''(\omega)$ reflects the density of states (DOS).  It is then easy to check that the jump in $\chi''(\omega)$ arises from the excitation of magnons at ${\bf q} = (0, \pi)$ with energy $\omega_{\bf q} = 2J-4J^\prime$. 
Similarly, the peak in $\chi''(\omega)$ is determined by the flat maximum of the dispersion along the $(0,\pi)$ to $(\pi,\pi)$ direction with energy 
\begin{equation}
\omega_{peak}=2J \sqrt{ { J-2J^\prime \over J+2J^\prime } } \ .
\end{equation}
However, a fit of $\chi''(\omega)$ to the experimental data (Fig.~\ref{chidp_om}) shows that the accessible energies are still about $75$meV below any of these pronounced features, which makes it difficult to extract a precise value for $J^\prime$.

We now turn to the analysis of the two-layer systems. The only change in our starting point  Eq.(\ref{hub}) is the introduction of a hopping term $t_\perp$ between the planes. Diagonalizing the Hamiltonian as before, one obtains two valence and two conduction bands, whose dispersion is given by Eq.(\ref{f_disp}) with $\epsilon^{-}_{\bf k}$ replaced by  $\epsilon^{-}_{\bf k}\pm t_\perp$. The subsequent calculation of the transverse susceptibilities is performed along the same lines as before, and has been reported earlier for $t^\prime = 0$ \cite{Mor96}. Due to the coupling of the planes, one obtains 
a finite out-of-plane susceptibility with the two spins in Eq.(\ref{susc}) on 
different layers.  The final result for the in-plane and out-of-plane susceptibilities in RPA approximation is given by
\begin{eqnarray}
& &\chi^{+-}_{1,\alpha}({\bf q},{\bf q},\omega) = - { J \Big[ 1 - \nu_{\bf q} - {J^\prime \over J} ( 1-cos(q_x)cos(q_y) )\Big] \over \omega^2 - \omega_{opt}^2({\bf q}) +i \delta }  \nonumber \\
& & \hspace{0.5cm} \mp { J \Big[ 1 - \nu_{\bf q} +{J_\perp \over 2J}- {J^\prime \over J} ( 1-cos(q_x)cos(q_y)) \Big] \over \omega^2 - \omega^2_{ac}({\bf q}) + i\delta }, \nonumber \\
& & \chi^{+-}_{1,\alpha}({\bf q},{\bf q+Q},\omega) = -{\omega \over 2} \Bigg[ 
{ 1 \over \omega^2 - \omega^2_{opt}({\bf q}) + i\delta}  \nonumber \\
& & \hspace{1.5cm} \pm { 1 \over \omega^2 - \omega^2_{ac}({\bf q}) + i\delta} \Bigg] \ ,
\end{eqnarray}
where $\alpha=1,2$ (upper and lower sign), denotes the in-plane and out-of-plane susceptibilities, respectively. Note that $\chi({\bf q}, {\bf q+Q}, \omega)$ changes sign under the exchange of layers. The optical and acoustic magnon modes are doubly degenerate with dispersion
\begin{eqnarray}
\omega_{ac,opt}({\bf q})&=& 2J \Bigg\{ \Big[ 1 \mp \nu_{\bf q} + {J_\perp \over 2J}- {J^\prime \over J} ( 1-cos(q_x)cos(q_y)) \Big] \nonumber \\
& & \hspace{-0.5cm} \times \Big[ 1 \pm \nu_{\bf q} - {J^\prime \over J} ( 1-cos(q_x)cos(q_y)) \Big]
\Bigg\}^{1/2}
\label{disp2}
\end{eqnarray}
where the upper (lower) sign corresponds to the acoustic (optical) mode.
The gap $\Delta=2\sqrt{J J_\perp}$ of the optical (or even) mode at $Q=(\pi,\pi)$ is not changed by $J^\prime$, however, the spin-wave velocity of the acoustic (or odd) mode is changed and now given by
\begin{equation}
c_{sw} = \sqrt{2} J \sqrt{1-2{J^\prime \over J}} \sqrt{1+ {J_\perp \over 4J} } \ .
\label{csw2}
\end{equation}
It is  important to note that since both $\Delta$ and $c_{sw}$ are experimentally measured, $J^\prime$ does not only lead to a renormalization of $J$, but also of  $J_\perp$. Using Eq.(\ref{csw2}) one obtains (in the limit of small $J_\perp$)
\begin{equation}
J_\perp = {\sqrt{2} \over c_{sw} } \sqrt{1-2 {J^\prime \over J}} \Big( { \Delta \over 2} \Big)^2
\end{equation}
Using $t^\prime = -0.45t$, we obtain from a fit of Eq.(\ref{disp2}) to the experimental data \cite{Hay97}, $J=163$ meV, $J^\prime = 33$ meV, and $J_\perp = 8.4$ meV. The value for $J$ is about $30\%$ larger, and the value for $J_\perp$ about $25\%$ smaller than the ones reported in Ref. \cite{Hay97}. It is interesting that the different values for $t^\prime$ in La$_2$CuO$_4$ and YBa$_2$Cu$_3$O$_{6.15}$  lead to almost identical values for the nearest-neighbor exchange constants, in contrast to the results in Ref.~\cite{Hay97}.  

\begin{figure} [t]
\begin{center}
\leavevmode
\epsfxsize=7.5cm
\epsffile{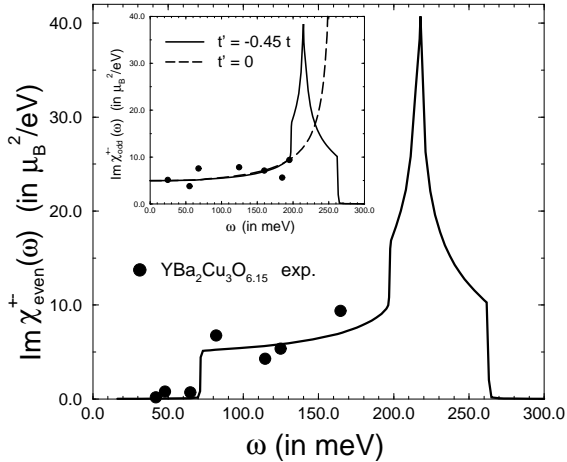}
\end{center}
\caption{The local susceptibility as a function of $\omega$ for the even channel and (see inset) the odd channel. The solid line corresponds to $t^\prime/t=-0.45$, whereas the dashed line shows the result for $t^\prime/t=0$. The experimental data are taken from Ref.\protect\cite{Hay97}.}
\label{chidp_om2}
\end{figure}

In Fig. \ref{chidp_om2} we present a fit of the local susceptibility in the even and odd channel, $\chi''_{even,odd}(\omega) = \chi''_{11} \pm \chi''_{12}$, to the experimental  data.
The fit in both channels is quite good, though the scattering of the data is rather large. A finite $t^\prime$ again transforms the divergence of $\chi''(\omega)$ at $\omega_{max}$ into a maximum. 
Note, that the two maxima in $\chi''_{even,odd}$ do not occur at the same frequency, but are shifted by a finite amount $\Delta \omega = 2 J_\perp (J^\prime/J)/\sqrt{1-(2 J^\prime/J)^2}\approx 3$meV. 
As before in the case of a single-layer, one can use the position of the jump and the maximum together with the value for $c_{sw}$ to extract $J^\prime$.  However, as a comparison of $\chi''$ for $J^\prime=0$ (dashed line) shows, the experimental data have to be extended to higher frequencies before one can extract $J^\prime$. 

Furthermore, using the data in Figs.~\ref{chidp_om} and \ref{chidp_om2}  we compared the local in-plane susceptibility $\chi''_{11}(\omega)$ in  YBa$_2$Cu$_3$O$_{6.15}$ and La$_2$CuO$_4$. We obtain that for $\omega<\Delta=72$ meV,  $\chi''_{11}(\omega)$ in  YBa$_2$Cu$_3$O$_{6.15}$ is only half as large as in  La$_2$CuO$_4$, which is expected since the optical mode is gaped out. For $72$ meV $<\omega $, the optical magnons in  YBa$_2$Cu$_3$O$_{6.15}$ 
can be excited and $\chi''_{11}$ in both materials is approximately the same.

Finally, we consider quantum corrections in the presence of $J^\prime$. A convenient way to calculate these corrections is to perform a systematic $1/S$-expansion in the Heisenberg model, as was done for the case $J^\prime=0$ in Ref.\cite{Iga92}. Considering a finite $J^\prime$ we performed calculations analogous to \cite{Iga92}. To lowest order in $1/S$, i.e., on the mean-field level, we recover as expected Eqs.(\ref{chiQ}) and (\ref{disp}). The next order $1/S$ corrections, which appear due to the normal ordering of quartic operator terms yield the following renormalizations (for $J^\prime/J=0.09$) 
\begin{eqnarray}
\tilde{J}&=&\Big(1+ {0.157\over 2S} \Big) \; J \, , \quad \tilde{J^\prime}=\Big(1- {0.123\over 2S}\Big) J^\prime  \ , \nonumber \\
\tilde{c}_{sw}&=& \Big(1+ {0.188\over 2S} \Big) c_{sw} \; , \quad Z_d = 1- { 0.454 \over 2S} \ .
\end{eqnarray}
Comparing these results with the case $J^\prime=0$, we find that the spin-wave velocity is only slightly increased whereas $Z_d$ is decreased by about $10\%$.
Can these results explain the strong renormalization of $\chi''$ observed experimentally ? A fit of $\chi''(0)=Z_d g^2 \mu_B^2/\sqrt{2} \tilde{c}_{sw}$ to the experimental data \cite{Hay97} showed, that the overall renormalization $Z_d^{exp}=0.36$ of the dynamical spin susceptibility in La$_2$CuO$_4$ was much larger than the theoretical value $Z_d=0.607$ obtained in the case $J^\prime=0$ \cite{Iga92}. For $J^\prime \not =0$ the overall renormalization of $\chi''$ is given by a classical contribution, $1/\sqrt{1-2\tilde{J}^\prime/\tilde{J} }$, and the quantum corrections $Z_d$, which means that $Z_d^{exp}$ should be compared with $Z^\prime=Z_d/\sqrt{1-2\tilde{J}^\prime/\tilde{J} }$. Here we use the  renormalized $\tilde{J}, \tilde{J}^\prime$ and $\tilde{c}_{sw}$, since only these are experimentally measurable. With $\tilde{J}^\prime/\tilde{J}=0.09$, which neglecting small corrections to order $(1/S)^2$ \cite{Iga92} implies $J^\prime/J=0.126$, we obtain $Z^\prime = 0.57$.  
Though this value is only about $6\%$ smaller than the one in Ref.\cite{Iga92}, it shows that a finite $J^\prime$ leads to an improved agreement with the experimental data. 

In conclusion we have provided a detailed study of the effects of $t^\prime$, and consequently $J^\prime$, on the spin dynamics in single and double-layer systems. We have shown that $J^\prime$ changes the value for $J$ and $J_\perp$ extracted from the experimental data, yielding two almost identical  values for $J$ in YBa$_2$Cu$_3$O$_{6.15}$ and La$_2$CuO$_4$.  We demonstrated that the changes due to $J^\prime$ in the high frequency structure of $\chi''(\omega)$ provide the possibility to extract  values for the exchange constants. Precise results, however, will require a higher accuracy of the INS data and their extension to higher frequencies. Furthermore, we obtained that $J^\prime$ leads to stronger quantum corrections. Finally, it was recently pointed out that the $^{17}$O NMR signal in Ni doped YBa$_2$Cu$_3$O$_{6+x}$ is sensitive to the momentum structure of $\chi'({\bf q})$ \cite{Bob97}; an effect which might be used to detect changes in $\chi'(q)$ due to $J^\prime$. Further quantitative studies in this direction are clearly called for.

We would like to thank A.V. Chubukov, S. Hayden, D. Pines, A. Sandvik, J. Schmalian and C. Slichter for valuable discussions. This work has been supported in part by the Science and Technology Center for Superconductivity through NSF-grant DMR91-20000.

\end{document}